\documentclass[twocolumn,showpacs,showkeys,preprintnumbers,amsmath,amssymb]{revtex4}

\usepackage{graphicx}
\usepackage{dcolumn}
\usepackage{bm}

\begin{document}

\title{Transport in disordered two-dimensional topological insulator}
\author{G.M.Gusev,$^1$ Z.D.Kvon,$^2$ O.A.Shegai,$^{1,2}$ N.N.Mikhailov,$^2$
S.A.Dvoretsky,$^{2}$ and J. C. Portal$^{3,4,5}$}

\affiliation{$^1$Instituto de F\'{\i}sica da Universidade de S\~ao
Paulo, 135960-170, S\~ao Paulo, SP, Brazil}
\affiliation{$^2$Institute of Semiconductor Physics, Novosibirsk
630090, Russia} \affiliation{$^3$LNCMI-CNRS, UPR 3228, BP 166, 38042
Grenoble Cedex 9, France} \affiliation{$^4$INSA Toulouse, 31077
Toulouse Cedex 4, France} \affiliation{$^5$Institut Universitaire de
France, 75005 Paris, France}

\date{\today}
\begin{abstract}
We study experimentally the transport properties of "inverted" semiconductor $HgTe$-based quantum well,
which is related to the two-dimensional topological insulator, in diffusive transport regime.
 We perform nonlocal electrical measurements in the absence of the magnetic field
and observe large signal due to the edge states. It demonstrates,
that the edge states can propagate over long  distance $\sim 1 mm$,
and, therefore, there is no difference between local and non local
electrical measurements in topological insulator. In the presence of
the in-plane magnetic field we find strong decrease of the local
resistance and complete suppression of the nonlocal resistance. We
attribute this observation to the transition between topological
insulator and bulk metal induced by the in-plane magnetic field.


\end{abstract}

\maketitle

Recently, a novel class of the topological state of a quantum matter
has emerged, called topological insulator \cite{kane, bernevig, Qi,
konig}. The first famous example of the two dimensional topological
insulator (2DTI) is the integer quantum Hall effect (QHE) state on
the resistance minima \cite{prange}. Such system represents the 2D
electron gas in the presence of the strong perpendicular magnetic
field, when all current is carried by the edge states, while
electrons in the bulk region are localized. The number of the edge
states are determined by the Landau level filling factor $\nu$. The
edge states are unidirectional because of the breaking of the
time-reversal (TR) symmetry due to the magnetic field and robust
against impurity scattering. Note, that disorder plays the crucial
role in establishment of the Hall plateau in the finite region of
the magnetic field or electron density.

Another class of 2DTI is the quantum spin Hall effect state, which
can be realized in 2D system with strong spin-orbit interaction in
the absence of the magnetic field \cite{kane, bernevig, Qi, konig}.
It has been shown that an "inverted" semiconductor HgTeCdTe quantum
well \cite{bernevig,konig}, possess the insulating phase having the
gap in the bulk electron spectrum and a single pair of
counterpropagating or helical edge states for the two opposite spin
polarizations. Remarkable consequence of the presence of the edge
states in ballistic 2DTI is a quantized longitudinal resistance
$R=h/2e^2$, which has been observed in micrometer scale Hall bars
\cite{konig}. The edge states are expected to be insensitive to
weak, nonmagnetic impurity scattering due to destructive
interference between two backscattering paths of the helical edge
states ( see review \cite{hasan}). It is worth noting that the
backscattering does not destroy the edge states, therefore any 2DTI
sample represents almost ideal natural one dimensional (1D) wire
localized near the sample edge.

In this letter we present an experimental study of the transport
properties of "inverted" $HgTe$-based quantum well which possesses a
pair of counter-propagating or helical edge states strongly mixed by
spin flip scattering. When the Fermi level passes the bulk
insulating gap, the devices revel unusual behaviour, which is
characterized by high resistance $R>>h/e^2$ with metallic
temperature dependence. Since the scattering between helical edge
states is topologically protected due to time reversal symmetry, we
attribute high level of the resistance to spin flip scattering
induced by spin-orbital interaction, or two-particle backscattering,
which breaks TR symmetry. Therefore, the edge state at the periphery
of the disordered TI constitutes the single one-dimensional wire. We
perform nonlocal electrical measurements and dramatically
demonstrate that edge-state transport truly exists over macroscopic
distance of $\sim 1 mm$ in the absence of the magnetic field.  The
in-plane magnetic field strongly suppresses local magnetoresistance
an completely destroys resistance measured in nonlocal
configuration. This observation is attributed to in-plane magnetic
field induced transition from 2D TI state to bulk metal state.

The $Cd_{0.65}Hg_{0.35}Te/HgTe/Cd_{0.65}Hg_{0.35}Te$ quantum wells
with the (013) surface orientation and a different thickness of 8
and 8.3 nm were prepared by molecular beam epitaxy. A detailed
description of the sample structure has been given in
\cite{kvon,gusev}. The schematic view of a typical micro-fabricated
device is shown in Figures 1a,b. Device consists of 3 narrow ($5 \mu
m$ of width) channels (segments) with different lengths ($8, 20 , 8
\mu m$ consequently), and 8 side branches (probes).  The ohmic
contacts to the two-dimensional gas were formed by the in-burning of
indium. To prepare the gate, a dielectric layer containing 100 nm
$SiO_{2}$ and 200 nm $Si_{3}Ni_{4}$ was first grown on the structure
using the plasmochemical method. Then, the TiAu gate was deposited.
Dimension of the gate is macroscopic ($1000\times500 \mu m^{2}$),
which is crucial for creating of the long length edge states. The
density variation with gate voltage was $1.09\times 10^{15}
m^{-2}V^{-1}$. The magnetotransport measurements in the described
structures were performed in the temperature range 0.050-4.1 K and
in magnetic fields up to 15 T using a standard four point circuit
with a 2-3 Hz ac current of 0.1-1 nA through the sample, which is
sufficiently low to avoid the overheating effects.
\begin{figure}[ht!]
\includegraphics[width=9cm,clip=]{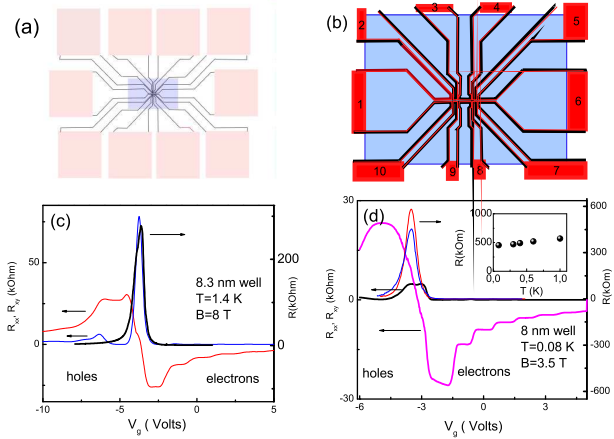}
\caption{\label{fig.1}(Color online) Schematic view (a,b) of the
sample. Red lines are edge states localized at the periphery of the
sample under TiAu metallic gate (blue area)in TI regime. Red areas-
are region with n-type HgTe. Diagonal $R_{xx}$ and Hall $R_{xy}$
resistances as a function of the gate voltage at zero and fixed
magnetic field for 8.3 nm (c) and 8.0 (d) nm HgTe quantum well.
Insert shows temperature dependence of the peak at zero magnetic
field. Two traces at zero field are shown in figure 1d (red T=1 K,
blue T=0.08K).   }
\end{figure}

Figures 1c,d show the data in zero magnetic field and standard
quantum Hall effect measurements in 8.3 nm (c) and 8.0 nm (d) HgTe
wells, which indicate high quality of our sample. The measured
mobility of electrons is approximately $\mu_{n}=250000 cm^{2}/Vs$
and mobility of  holes $\mu_{p}\approx25000 cm^{2}/Vs$, which is
comparable to that found in HgTe well of wider width \cite{kvon,
gusev}. Sweeping gate voltage $V_{g}$ from the positive to the
negative value than depopulates electrons states and populates hole
states. We may see it from Hall resistance behaviour: $R_{xy}$
changes smoothly through zero from its negative  quantized value on
the electron side to a positive quantized value on the hole side,
whereas $R_{xx}$ moves from a zero on the electron side through a
maximum at the bulk energy gap to another zero-minimum on the hole
side. Note that the resistance at maxima sharply decreases with
magnetic field.
\begin{figure}[ht!]
\includegraphics[width=9cm,clip=]{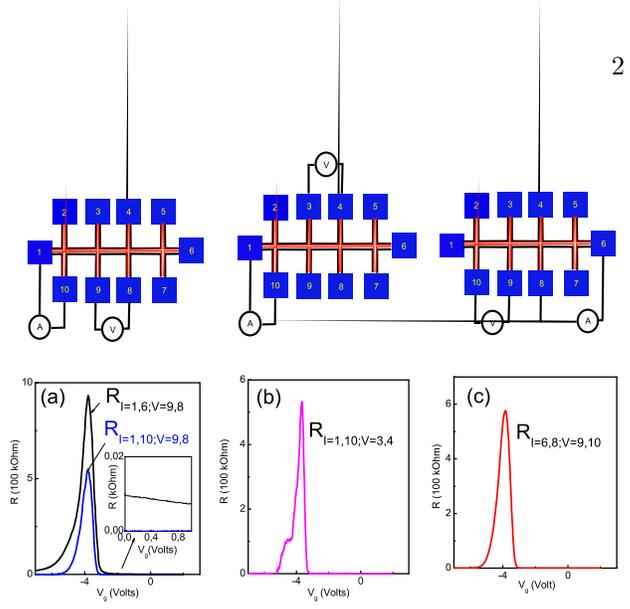}
\caption{\label{fig.2}(Color online) Local (a) and nonlocal (a,b,c)
resistances at zero magnetic field  as a function of the gate
voltage for different configurations, which are shown on the top
panel. Red lines represent 1D wire along sample edge. Insert to
Fig.2a shows the local and nonlocal resistance in the region with
bulk dissipative transport. T=80 mK}
\end{figure}

In the following we will concentrate on the electronic behavior
around the bulk energy gap in zero magnetic fields and at low
temperatures. It is expected that the transport properties in this
region are determined by pair of counterpropagating or helical edge
states states for the two opposite spin polarizations, and
resistance is quantized at value $h/2e^2$. However, we find finite
resistance $R\approx300 kOhm$ (8.3 nm well), which is significantly
higher than $h/2e^2$. This agrees with previous results obtained in
samples with a length of $L = 20 \mu m$, which has been attributed
to inelastic scattering between helical edge states \cite{konig2}.
Nonuniversality of the resistance in the presence of the spin
dephasing has been predicted by various theoretical models
\cite{jiang,strom}. Inelastic mechanism, however, disagrees with
absence of the temperature dependence of the resistance peak.  The
origin of the resistance is determined by the behaviour of the
contacts regions, similar to the conventional quantum Hall effect.
Contacts are assumed to be a thermal reservoirs \cite{buttiker},
when full mixing of electron spin states can occur. Note, that in
contrast to the standard QHE, when  mixing of edge states occurs
within metallic Ohmic contacts, in our samples potential mixing is
provided by 2D electron gas  in the region outside of the metallic
gate due to finite bulk conductivity. Fig. 1b illustrates such
geometry. We may see that the length of the edge states are
determined by the perimeter of the sample part covered by metallic
gate (mostly side branches) rather that the length of the
microbridge. We calculate the length of  1D wires between probe
contacts and find that it exceeds 1.0 mm. For such macroscopic
distance it is naturally to expect strong backscattering between
states with two opposite spin polarizations due to spin flip
scattering or scattering by magnetic impurities.  Elastic magnetic
scattering is unlikely to explain our data, because the background
magnetic impurity content of MBE machine is very low. The high
resistance $R>>h/e^2$ can be attributed to the spin flip scattering
induced by spin-orbital interaction, somewhat similar to transition
between spin split states in the quantum Hall effect regime,
considered in the model \cite{khaetskii}. Such elastic
backscattering should lead, however, to localization, and
T-dependent resistance. This is contrary to our observations.  The
alternative model based on two-particle excitation
\cite{strom,xu,wu} leads to suppression of localization, since the
inelastic dephasing time becomes comparable with backscattering
time, which can cancel out the  temperature effect.

Note that the 2DTI in zero magnetic field is equivalent to the QHE
state at $\nu=0$ near charge neutrality point in graphene considered
in the model \cite {abanin}. Within this model transport in graphene
at strong magnetic field is dominated by the pair of gapless edge
excitations, which however, have magnetic nature.  In the absence of
transport through the bulk, if both edges carry the same current,
the resistance is $R=\frac{h}{2e^2}(1+\gamma L)$, where
$\gamma^{-1}$ is the mean free path for 1D backscattering, L is the
length of 1D channel between contacts, or thermal reservoirs, which
in our case are related to 2D electron gas. Comparing this model,
which applies equally well to the transport in 2DTI in zero magnetic
field,  with our results we obtain $\gamma^{-1}=400\mu m$.
Estimation of the two-particle backscattering predicts corresponding
length of the order of few meters \cite{ostrovsky}. More theoretical
work is desirable in order to understand spin flip dephasing
mechanism in TI. We measured several samples from different wafers
and found that peak resistance is varied from 300 kOhm to 2 MOhm,
which likely stems from variation of the disorder parameters. The
application of the current between any pair of the probes creates
net current along the sample edge, and can be detected by any other
pair of the voltage probes. In general, there is no difference
between local and non local electrical measurements in topological
insulator. Fig.2 shows several traces for local and nonlocal
resistance taken at different configurations for 8 nm HgTe well. The
data demonstrate that the nonlocal resistance is negligibly small in
the region with homogeneous 2D electron and hole gas, outside of the
TI peak, as expected for classical dissipative transport (see insert
to figure 2a). The nonlocal resistance in TI regime has comparable
amplitude and qualitatively the same position and width of the peak.
The occurrence of the nonlocal resistance of comparable amplitude
 implies that potential difference extend over macroscopic distance
 $\sim 1 mm$ away from dissipative bulk current path. It is only
 possible, if the bulk conductivity is completely suppressed and
 conductivity of the TI sample is determined by edge states. To
 analyze nonlocal resistance value we employ simple model which describes
 sample resistance as combination of resistors along sample edge,
 assuming that resistance value is proportional to the length of the
 1D channel (Figure 2). For
 realistic configuration of the sample, shown in figure 1, we
 estimate this
 ratio $\sim 0.45$, which roughly agrees with experimental value.
Note that in ballistic case the nonlocal transport indeed has been
observed in mesoscopic samples \cite{roth}, when nonlocal resistance
have been obtained within the general Landauer-B\"{u}ttiker
formalism \cite{buttiker}.

\begin{figure}[ht!]
\includegraphics[width=9cm,clip=]{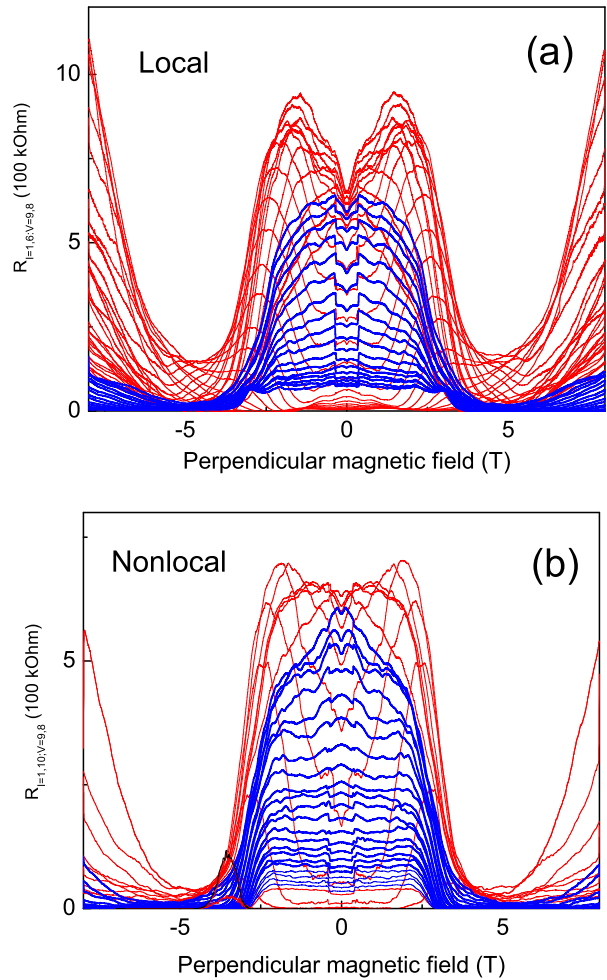}
\caption{\label{fig.3}(Color online) (a) Local resistance
 as a function of the perpendicular magnetic field for different gate voltages.
 Red traces a taken for electronic part of TI peak, $V_{g}$=-2.5- (-3.8)Volts, step 0.05 Volts.
 Blue traces are taken for hole part of TI peak, $V_{g}$=-3.8- (-4.5)Volts, step 0.05 Volts.
(b) Nonlocal resistance
 as a function of the perpendicular magnetic field for different gate voltages.
 Red traces a taken for electronic part of TI peak, $V_{g}$=-2.5 - (-3.8)Volts, step 0.1 Volts.
 Blue traces are taken for hole part of TI peak, $V_{g}$=-3.8 - (-4.5)Volts, step 0.05 Volts.
 T=80 mK.}
\end{figure}

\begin{figure}[ht!]
\includegraphics[width=9cm,clip=]{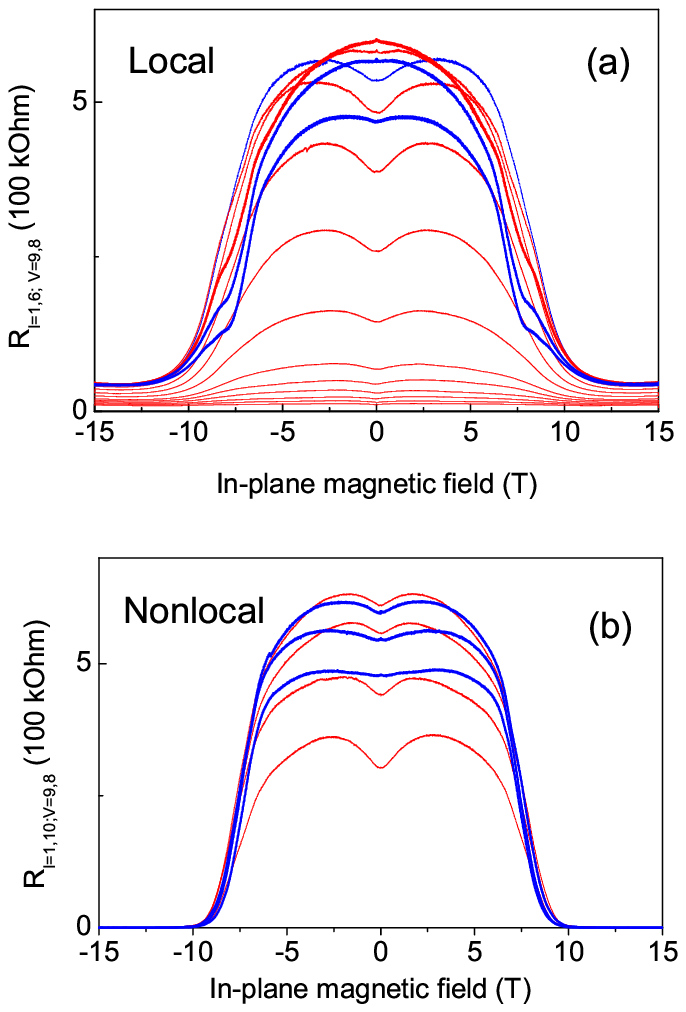}
\caption{\label{fig.4}(Color online) (a) Local resistance
 as a function of in-plane magnetic field for different gate voltages.
  Red traces a taken for electronic part of TI peak, $V_{g}$=-2.5- (-3.7)Volts, step 0.05 Volts.
 Blue traces are taken for hole part of TI peak, $V_{g}$=-3.8- (-4.0)Volts, step 0.1 Volts.
(b) Nonocal resistance
 as a function of in-plane magnetic field for different gate voltages.
  Red traces a taken for electronic part of TI peak, $V_{g}$=-3.5- (-3.8)Volts, step 0.1 Volts.
 Blue traces are taken for hole part of TI peak, $V_{g}$=-3.9- (-4.1)Volts, step 0.1 Volts.
 T=80 mK.}
\end{figure}

 Application of the strong perpendicular, and
in-plane magnetic field  would be important to establish the physics
of topological insulators, and one may ask if they can also exhibit
other interesting phenomena. Previous experiments in ballistic
structures \cite{konig} demonstrated sharp spike at B=0 and decrease
of the conductance with increasing of perpendicular magnetic field
$B_{\bot}$. No magnetoresistance has been observed in the presence
of the in-plane magnetic field $B_{\|}$. This observation has been
attributed to the opening of the gap between edge states due to
Zeeman splitting. Alternative scenario has been suggested in paper
\cite{maciejko}, when orbital effects and effects of disorder have
been included. In disordered TI we find different behaviour. Figure
3 shows the magnetoresistance in local and nonlocal configurations
for different gate voltages, when the Fermi level passes from
electron to hole side of the peak in figure 2. It is worth noting
remarkable similarity between these two sets of curves. It confirms
again that the transport in TI occurs only at the periphery of the
sample. The magnetoresistance exhibits several features with
magnetic field. First, sample reveals the spike a low magnetic
field, which, however, has a smaller magnitude that one in ballistic
devices, and shows remarkable evolution with gate voltage. Second,
magnetorsistance exhibits several jumps and sawtooth oscillations
with magnetic field, which are very different from conventional
universal conductance fluctuations in 1D wires \cite{stone}. Third,
we may see giant magnetooscillations with magnitude $\sim
 200-400 kOhm $, which is also very unusual in 1D wires. Finally,
 strong decrease of the resistance at $B>4$T  is observed, moreover, in electronic side of the peak,
 resistance approaches zero, which  corresponds $\nu=1$ minima in the
 quantum Hall effect regime. Such rich magnetorsistance behavior
 requires more detailed analysis and comparison with theory, and indeed
it will be done in forthcoming publications.

The Figure 4 shows the evolution of the local and nonlocal
magnetoresistance with gate voltage in the presence of the in-plane
magnetic field. Again we may see strong similarity between these two
sets of magetoresistance curves. In contrast to previous
measurements in ballistic devices \cite{konig2} we observe large
positive magnetoresistance within 2 T. In magnetic field $B_{\|}> 6
T$ both resistances dramatically drop with increase of the in-plane
field. Examination of traces in Fig 4 in magnetic field higher than
10 T immediately reveals drastic difference between $R_{xx}$ and
nonlocal resistance: the nonlocal resistance data with negligible
small value in this region corresponds to large (10-50 kOhm) value
of $R_{xx}$. Note, that in perpendicular magnetic field both
resistances have comparable values in strong B ( figure 3). Thus,
the external parallel magnetic field strongly suppresses $R_{xx}$
and completely destroys nonlocal resistance. The nonlocal resistance
is very small in the presence of the bulk dissipative transport.
Therefore it is naturally suggest, that in-plane magnetic field
destroys the edge state transport and at the same time induces
electronic states in the bulk. Alternatively we may suggest that
in-plane field induces unidirectional edge states similar to QHE
state in perpendicular magnetic field. These states indeed are
dissipationless, and voltage drop between probe contacts is zero.
However, within this scenario, local $R_{xx}$ should be zero too,
which disagrees with our observation (see Fig.4a). Therefore our
data demonstrate the existence of the in-plane magnetic field
induced transition between 2DTI state and bulk metal state. More
experimental work and theoretical effort has to be done in order to
understand the origin of these emergent phenomena.

In summary, we report on the observation of disordered 2DTI state in
HgTe quantum wells in zero magnetic field, and the emergence of
conductive bulk state in the presence of the in-plane magnetic
field. We demonstrate the similarity between local and nonlocal
resistance measurements in zero and perpendicular magnetic field in
TI regime. These data give the evidence  that in realistic samples
edge-state transport truly exists over macroscopic distance of $\sim
1 mm$ in the absence of the magnetic field.

We thank O.E.Raichev and I.V.Gorny for helpful discussions. A
financial support of this work by FAPESP, CNPq (Brazilian agencies),
USP-COFECUB (Uc Ph 109/08), FAPESP-CNRS, RFBI (09-02-00467a and
ofi-m) and RAS programs "Fundamental researches in nanotechnology
and nanomaterials" and "Condensed matter quantum physics" is
acknowledged.

\end{document}